\documentclass[prb,twocolumn,preprintnumbers,amsmath,amssymb]{revtex4}
\usepackage{graphicx}
\usepackage{epsfig}
\usepackage{amsmath,amscd}
\usepackage{psfrag}

\def\eqref#1{Eq.~\ref{#1}}
\def\figref#1{Fig.~\ref{#1}}

\def\Hsh{H_{s}}
\def\ush{u_{s}}
\def\Ysh{z(\ush)}
\def\Deltash{\Delta_{s}}
\def\TcCleanStatics{T_{c}}
\def\DoS{\nu}

\def\td{\textup{d}}

\def\romega{\epsilon}
\def\omegan{\omega_n}
\def\gapsh{\romega_s}

\def\avf{\langle f \rangle}
\def\avg{\langle g \rangle}

\def\avgcos{\langle g\cos\theta \rangle}
\def\taum{\tau_-}
\def\taup{\tau_+}
\def\malpha{\alpha_m}
\def\mtau{\tau_m}

\def\im{~\rm Im~}

\begin{document}

\title{Effect of Impurities on the Superheating field of Type II superconductors}
\author{F. Pei-Jen Lin$^{1,2}$}\email{fareh.lin@gmail.com; flin@odu.edu}
\author{A. Gurevich$^1$}\email{gurevich@odu.edu}
\affiliation{$^1$Department of Physics, Old Dominion University,
Norfolk, Virgina 23529, USA}
\affiliation{$^2$Argonne National Laboratory, Argonne, Illinois 60439, USA}
\begin{abstract}

We consider the effect of nonmagnetic and magnetic impurities on the superheating field $H_s$ in a type-II superconductor. We solved
the Eilenberger equations, which take into account the nonlinear pairbreaking of Meissner screening currents, and calculated $H_s(T)$ for arbitrary temperatures and impurity concentrations in a single-band s-wave superconductor with a large Ginzburg-Landau parameter. At low temperatures nonmagnetic impurities suppress a weak maximum in $H_s(T)$ which has been predicted for the clean limit, resulting instead in a maximum of $H_s$ as a function of impurity concentration in a moderately clean limit. It is shown that nonmagnetic impurities weakly affect $H_s$ even in the dirty limit, while magnetic impurities suppress both $H_s$ and the critical temperature $T_c$. The density of quasiparticles states $N(\epsilon)$ is strongly affected by an interplay of impurity scattering and current pairbreaking. We show that a clean superconductor at $H=H_s$ is in a gapless state, but a quasiparticle gap $\epsilon_g$ in $N(\epsilon)$ at $H=H_s$ appears as the concentration of nonmagnetic impurities increases. As the nonmagnetic scattering rate $\alpha$ increases above $\alpha_c=0.36$, the quasiparticle gap $\epsilon_g(\alpha)$ at $H=H_s$ increases, approaching $\epsilon_g\approx 0.32\Delta_0$ in the dirty limit $\alpha\gg 1$, where $\Delta_0$ is the superconducting gap parameter at zero field. The effects of impurities on $H_s$ can be essential for the nonlinear surface resistance and superconductivity breakdown by strong RF fields.

\end{abstract}

\pacs{74.25.-q, 74.25.Ha, 74.25.Op, 74.78.Na}

\maketitle

\section{introduction}
Type -II superconductors are in the Meissner state if the applied magnetic field $H$ is smaller than the lower critical magnetic field $H_{c1}$ above which the Gibbs free energy of a vortex becomes negative. However, the Meissner state can remain metastable at higher magnetic fields, $H>H_{c1}$ up to the superheating field $H_s$ at which the Bean-Livingston surface barrier \cite{Bean64} for penetration of vortices disappears and the Meissner screening currents at the surface become unstable with respect to small perturbations of the order parameter. The field $H_s$ is therefore the maximum magnetic field at which a type-II superconductor can remain in a true non-dissipative state not altered by dissipative motion of vortices. In addition to the fundamental interest in what limits $H_s$ in superconductors, there is a strong interest in the physics of $H_s$ promoted by recent advances in superconducting Nb cavities for particle accelerators in which the quality factors $Q\sim 10^{10}-10^{11}$ at 2K and 2GHz were observed up to the surface RF fields close to the thermodynamic critical field $H_c\simeq 200$ mT of Nb. \cite{cavity} Here the RF frequency is well below the gap frequency $\Delta_0/h\simeq 400$ GHz for Nb where $\Delta_0$ is the superconducting gap, so the current pairbreaking in the cavities at low temperatures may occur under quasi-static conditions.

Calculations of the superheating field $H_s$ and the depairing current density $J_c$ have a long history starting from the pioneering works by Ginzburg \cite{GL}, de-Gennes \cite{deGennes65} and Matricon and Saint-James \cite{Matricon67} who used the Ginzburg-Landau (GL) theory. It was shown that as the applied field $H$ reaches $H_s$, the Meissner screening current density at the surface becomes of the order of $J_c\simeq c H_c/4\pi\lambda$, which makes the superconducting state unstable. The value of $H_s$ in the limits of large and small GL parameter $\kappa$ is given by \cite{Kramer68,Fink69,Chapman95,Dolgert96,Parr,corn}
\begin{gather}
\Hsh= (\sqrt{5}/3)H_c, \qquad \kappa\gg 1,
\label{h1} \\
\Hsh=2^{-1/4}\kappa^{-1/2}H_c, \qquad \kappa \ll 1.
\label{h2}
\end{gather}
Here the factor $\sqrt{5}/3$ in Eq. (\ref{h1}) reflects the suppression
of the order parameter by the Meissner currents in the local limit $\kappa \gg 1$,
while the enhancement of $H_s$ by the factor $\kappa^{-1/2}$ for $\kappa\ll 1$
results from the proximity effect reduction of the Meissner pairbreaking localized in a narrow
surface layer of thickness $\lambda \ll \xi$, where $\lambda$ is the London penetration depth
and $\xi$ is the coherence length. The extensive calculations of $H_s$ based on the GL equations
\cite{Kramer68,Fink69,Chapman95,Dolgert96,Parr,corn} have shown that, for $\kappa \gg 1$, the Meissner state becomes absolutely unstable
with respect to small 2D perturbations of current and order parameter with the wavelength $\sim (\xi^3\lambda)^{1/4}$ along the surface and decaying
over the length $\sim \sqrt{\lambda\xi}$ perpendicular to the surface. Such perturbations describe the initial stage of penetration of vortex rows with the period $\sim (\phi_0/H_s)^{1/2}$ corresponding to the equilibrium vortex lattice at the field $H=H_s$.

Unlike $H_s(T)$ in the GL region near $T_c$, the behavior of $H_s(T)$ at low temperatures is not well understood, not least because the calculation of $H_s(T)$
requires solving the nonlinear Gor'kov or Eilenberger equations \cite{Eilenberger68} which take into account pairbreaking effect of Meissner currents. Manifestations of pairbreaking effects in a clean superconductor at low temperatures can differ from the GL results, as was shown long ago by Parmenter \cite{Parmenter62} and Bardeen \cite{Bardeen62}. The first calculation of $H_s(T)$ for the entire temperature range $0<T<T_c$ in the clean limit and $\kappa\to \infty$ was done by Galaiko \cite{Galaiko66} who obtained $\Hsh=0.84H_c$ at $T\rightarrow0$, and $\Hsh=(\sqrt{5}/3)H_c = 0.745H_c$ at $T\rightarrow T_c$. Catelani and Sethna \cite{Catelani08} solved the Eilenberger equations to calculate the temperature dependence of $H_s(T)$ for $0<T<T_c$ in the clean limit for $\kappa\to\infty$ and found a maximum in $H_s(T)$ at low $T$. Such a non-monotonic temperature dependence of $H_s(T)$ shows that the behavior of $H_s(T)$ at low $T$ can hardly be extrapolated from the GL results near $T_c$.

The effect of impurities on $H_s(T)$ outside the GL region has not been addressed. This problem is of interest because the clean limit in s-wave superconductors at $T \ll T_c$ is a rather singular case: for $T=0$, the Meissner currents do not affect the superfluid density $n_s$ until the superfluid velocity $v_s = J/n_se$ reaches the critical value, $v_s = v_c =\Delta_{00}/p_F$ where $p_F$ is the Fermi momentum and $\Delta_{00}=\Delta_0(0)$ is the modulus of the order parameter $\Psi=\Delta\exp(i\varphi)$ at zero superfluid velocity and $T=0$. \cite{Parmenter62,Bardeen62} For $v_s>v_c$, the gap in the quasiparticle spectrum disappears and $n_s(v_s)$ rapidly drops to zero in a narrow region $v_c<v_s<1.08 v_c.$ \cite{Maki63,Makigapless} Thus, unlike the d-wave superconductors, the s-wave superconductors at $T\ll T_c$ do not exhibit the nonlinear Meissner effect caused by the dependence of $n_s(J)$ on the current density \cite{Yip,Dahm,Hirsch,Groll}, and the superheating field is reached at the superfluid velocity $v_s>v_c$, which corresponds to a {\it gapless} state \cite{Fulde65}. The latter has important consequences for the low-frequency $(\omega\ll\Delta_0)$ impedance of clean superconductors because at $H=\Hsh=0.84H_c$ the surface resistance $R_s$ becomes of the order of $R_s$ in the normal state,
unlike the exponentially small $R_s\sim \omega^2\exp(-\Delta_0/k_BT)$ in a fully gapped state at $H\ll \Hsh$. \cite{mb,rs}
The effect of the RF Meissner currents on the quasiparticle spectrum
can result in a strong dependence of the surface resistance on
the RF field amplitude with $H>TH_c/\Delta_0$. \cite{Gurevich06}

For $\kappa\gg 1$, the field $H_g$ at which the gap in the quasi-particle density of states $N(\epsilon)$ closes at $T=0$ can be calculated from the London equation $H_g=4\pi\lambda J_c/c$ where $J_c=nev_c$, $\lambda=(mc^2/4\pi n e^2)^{1/2}$, $n$ is the total electron density, $-e$ is the electron charge, $m$ is the band effective mass, and $c$ is the speed of light. Here the linear London screening of $H(x)=H_0e^{-x/\lambda}$ remains valid as long as the superfluid density is independent of $J$, that is, the Meissner current density $J=(c/4\pi)\partial H/\partial x$ is smaller than $J_c$.
This yields $H_g=c\Delta_{00}/e\lambda v_F$, which can be expressed in
terms of $H_c=\big(4\pi N(0)\big)^{1/2}\Delta_{00}$, where
$N(0)=m^2v_F/2\pi^2\hbar^3$ is the density of
states per one spin orientation for an isotropic parabolic band,
and $v_F=(3\pi^2n)^{1/3}\hbar/m$. Hence,
\begin{equation}
H_g=(2/3)^{1/2}H_c \approx 0.816 H_c.
\label{hg}
\end{equation}
Since $H_s=0.84H_c$, \cite{Galaiko66} the gapless state in the clean, large-$\kappa$
limit occurs in a narrow field range, $0.97H_s\lesssim H<H_s$.

The goal of this work is to address the pairbreaking effect of impurities on $H_s$. Indeed, while nonmagnetic impurities do not affect $T_c$, $\Delta_0$ and $H_c$ at zero superfluid velocity \cite{Anderson59,Abrikosov61}, these impurities become pairbreakers in the current-carrying state \cite{Makigapless}, which, for example, manifests itself in the nonlinear Meissner effect\cite{Yip,Dahm,Hirsch,Groll}. Thus, the extent to which nonmagnetic impurities would affect $H_s$ needs to be understood. The effects of impurities on the quasiparticle density of state $N(\epsilon)$ at $H=H_s$ and the conditions under which impurities can restore the gap in $N(\epsilon)$ at $H=H_s$ are important for the understanding of the nonlinear surface resistance at high fields and the limits of superconductivity breakdown under low frequency RF fields. We will also consider the superheating field in thin film multilayers consisting of alternating superconducting and dielectric layers thinner then $\lambda$. The parallel $H_{c1}$ of such multilayers is greatly enhanced, which enables probing the pairbreaking limits in parallel magnetic fields. It was suggested to use these multilayer coatings to increase the RF breakdown fields of superconducting cavities in particle accelerators. \cite{Gurevich06apl} The paper is organized as follows.

In Section II we solve the Eilenberger equations for a superconductor with uniform current
and analyzed the nonlinear dependencies of $\Delta$ and $J$ on the superfluid velocity in the presence of impurities.
In Section III we obtained equations for $H_s(T)$ in the entire temperature
range $0<T<T_c$ for $\kappa\gg 1$ and arbitrary concentrations of nonmagnetic and magnetic impurities. We solved the equations
for $H_s$ numerically and analyzed the dependencies of $H_s$ on the
magnetic and nonmagnetic scattering rates. The range of parameters in which $H_s$
can be optimized by varying the impurity concentration was found. In Section IV we consider the effect of impurities
on the quasiparticle density of states at $0<H<\Hsh$, particularly
the emergence of the gap in $N(\epsilon)$ as the concentration of
nonmagnetic impurities increases. Section V is devoted to the calculation of
nonlinear screening of strong DC field and $H_s$ in multilayers. In Section VI we discuss
our results and their implications for a nonlinear surface resistance
at high RF fields and superconductivity breakdown in cavities.

\section{Theory}\label{theory}

\subsection{Eilenberger Equations}
\label{Eeq}
In this work we use the Eilenberger equations for a single-band
s-wave superconductor with magnetic and nonmagnetic impurities:
\cite{Eilenberger68}
%-------------Eilenberger equation
\begin{equation}\label{EilenbergerEquation0}
(2\omegan+{\bf v} \cdot \mathcal D )f=2\Psi g+\frac1\taum g \avf-\frac1\taup f\avg,
\end{equation}
where $\Psi$ is the order parameter,
${\bf v}$ is the Fermi velocity,
$\omegan=\pi T(2n+1)$ are Matsubara frequencies, $T$ is the temperature,
$\mathcal D=\nabla +2\pi i {\bf A}/\phi_0$,
$\phi_0$ is the flux quantum, and $\bf A$ is the vector potential.
We use the units for which $\hbar= k_B=1$ unless stated otherwise.
The time constants $\taup$ and $\taum$ are defined by
%--------scattering rate
\begin{equation}
\frac1{\tau_{\pm}}=\frac1\tau\pm\frac1\mtau,
\label{tau}
\end{equation}
where $\tau$ is the electron scattering time on nonmagnetic impurities,
and $\mtau$ is the spin-flip scattering time on magnetic impurities.
The angular brackets $\langle \cdots \rangle=\int_{S_F} \td^2 k_F$ mean angular averaging over the Fermi surface.
The quasiclassical Green functions $f({\bf v},{\bf r},\omegan)$
and $g({\bf v},{\bf r},\omegan)$ are normalized by
%---------------normalization
\begin{equation}\label{normalization}
g^2+ff^{\dagger}=1.
\end{equation}
%-------------------------gap equation
where $f^\dagger({\bf v}, {\bf r}, \omegan)=f^*(-{\bf v}, {\bf r}, \omegan)$,
and the asterisk means complex conjugation.
The self-consistency equation for the superconducting order parameter $\Psi({\bf r})=\Delta\exp(i\varphi)$ is given by
\begin{equation}\label{gapeq0}
\Psi=2\pi T N(0) |V| \sum_{\omegan=0}^{\omega_D} \langle f \rangle,
\end{equation}
where $N(0)$ is the normal density of states per one spin at the Fermi surface,
$V$ is the BCS coupling constant and the cut-off frequency
$\omega_D$ is of the order of Debye frequency. Here $V$
can be expressed in terms of the critical temperature $\TcCleanStatics$ of a superconductor without nonmagnetic impurities:
$|V|^{-1}=N(0)\ln(2\omega_D \gamma/\pi \TcCleanStatics)$ where $\gamma=1.78$.
%---------------------------------vector potential equation
Eqs. (\ref{EilenbergerEquation0})-(\ref{normalization}) are supplemented by the Maxwell equation
for $\bf A$ and the supercurrent density ${\bf J}$:
\begin{gather}
\nabla\times\nabla\times {\bf A}=\frac{4\pi}c {\bf J}
\label{maxwellB} \\
{\bf J}=-4 \pi T e N(0) \im
\sum_{\omega_n=0}^{\infty} \langle {\emph g} {\bf v}\rangle.
\label{jeq0}
\end{gather}

\subsection{Geometry and assumptions}

To study the stability of the Meissner state with respect to small
perturbations, we consider a planar type-II superconductor occupying
the region $x>0$ with magnetic field ${\bf H}$ applied along the $z$-axis.
For a large-$\kappa$ superconductor, the previous calculations based on the GL
\cite{Fink69,Chapman95,Dolgert96,corn} and Eilenberger \cite{Catelani08} theories
have shown that the instability at $H=H_s$ is driven by coupled
fluctuations of $\delta\Psi(x,y)$ and $\delta{\bf J}(x,y)$ that
rapidly oscillate on the scale $\sim\xi\kappa^{1/4}$ parallel to
the surface, and decay on a longer length $\sim\xi\kappa^{1/2}$
perpendicular to the surface. Taking these inhomogeneous unstable
modes into account is essential for the calculation of small
corrections $\sim H_c\kappa^{-1/2}$ to $H_s$. However, the main
contribution to $H_s \sim H_c$ is determined by the condition that
the superfluid velocity $v(0)$ at the surface reaches the critical
pairbreaking value $v_c$ for which $J(v)$ is maximum and the
superconducting state becomes unstable. It is the same condition
that defines the uniform depairing current density \cite{Kupriyanov80}.

For uniform current flow, the solution of the Eilenberger equation can be sought in the form:
\begin{eqnarray}
\Psi({\bf r})&=&\Delta e^{iqy},
\label{ansatz1} \\
f({\bf r},\omegan,\theta)&=&f(\theta,\omegan) e^{iqy},
\label{ansatz2}
\end{eqnarray}
where $q$ is the wave vector of the condensate, where $\theta$ is the
angle between the Fermi velocity and the current,
and the amplitudes $\Delta$ and $f(\theta,\omegan)$ are independent
of the coordinates.
Equations (\ref{ansatz1}) and (\ref{ansatz2}) approximate the solution necessary for the calculation of $H_s$,
taking into account the difference in characteristic lengthscales in an extreme type-II superconductor with $\kappa\gg 1$. In this case $H_s$ is reached
as the current density at the surface becomes of the order of the depairing current density, which corresponds to the wavelength
$2\pi/q\simeq \xi$ \cite{Kupriyanov80}. Slow decrease of the Meissner
screening currents over the London penetration depth increases $H_s$ which is now limited by perturbations $\delta{\bf J}(x,y)$ and $\delta f(\omegan,\theta,x,y)$ oscillating on the scale
$2\pi/k\sim(\xi^3\lambda)^{1/4}$ along the $x-$axis and decaying over $\sim (\xi\lambda)^{1/2}$ along the $y-$axis \cite{Fink69,Chapman95,Dolgert96,corn}.
Thus, Eqs. (\ref{ansatz1}) and (\ref{ansatz2}) give an asymptotically
exact solution for the lower bound of $H_s$ in the limit of $\kappa\to\infty$ which will be addressed in this paper. The effect of London screening resulting in the finite-$k$ instability produces small corrections in $\Hsh=H_c (\sqrt{5}/3)(1+1/\sqrt{2\kappa})$ in the GL region \cite{Fink69,corn}.

\begin{figure}[ht]
\includegraphics[width=7cm,angle=0]{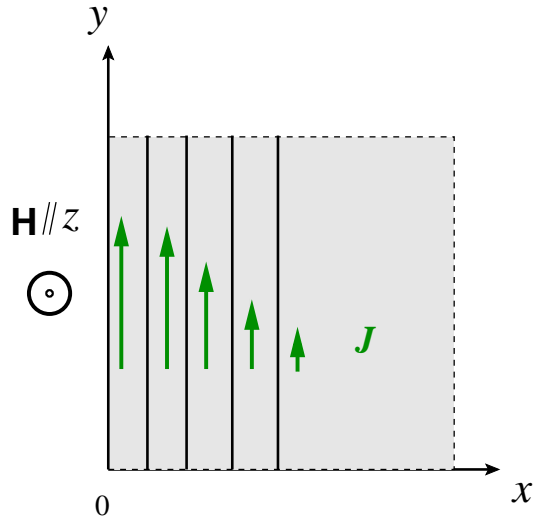}
    \caption{Schematic diagram of a multi-layered superconductor:
Superconducting layers (gray) separated by insulating layers (dark lines).
Applied magnetic field applied along the $z$-direction induces currents
flowing along $y$-direction.}
\label{fig1}
\end{figure}

Equations (\ref{ansatz1}) and (\ref{ansatz2}) can also be applicable for
a wider range of $\kappa$ in a multilayer system in which superconducting
layers are separated by thin dielectric layers as shown in \figref{fig1}.
For thin superconducting layers of thickness $d\ll\lambda$, the Meissner
current is nearly uniform across the film, while for $d<\sqrt{\lambda\xi}$,
the vortex instability is suppressed as the perpendicular components of ${\bf J}(x,y)$ cannot cross the dielectric layers.
\cite{Chapman95} In this case a uniform pairbreaking instability at $H=H_s$ develops, for which
equations (\ref{ansatz1}) and (\ref{ansatz2}) are exact solutions of the
Eilenberger equations. These solutions will be used to calculate
nonlinear screening of magnetic field in multilayers.

\subsection{Solution of the Eilenberger equations}

Substituting Eqs. (\ref{ansatz1}) and (\ref{ansatz2}) into Eq. (\ref{EilenbergerEquation0}) gives
the following algebraic equations for the amplitudes $g(\theta,\omegan)$ and $f(\theta,\omegan)$:
\begin{equation}\label{HomoEilenbergerEquation}
\frac g f= \frac{1}{2\Delta+\avf/\taum}\left( 2\omegan+\frac\avg\taup+iuv\cos\theta\right),
\end{equation}
where we assumed a spherical Fermi surface for which $\langle f \rangle =\frac{1}{2}\int_{0}^{\pi} f(\theta)\sin\theta d\theta$,
$\theta$ is the angle between the Fermi velocity and the direction of current. The gauge-invariant wave vector
$u=q+ 2\pi A_y/\phi_0$ is proportional to the superfluid velocity $v_s=\hbar u/2m$, where $m$ is the effective electron mass.
Solving Eqs. (\ref{normalization}) and (\ref{HomoEilenbergerEquation}) yields
%-------------------------------------------------------f and g in <f> and <g>
\begin{eqnarray}\label{newf}
f&=&\frac b{\sqrt{ b^2+\left(a+i\cos\theta \right)^2 }},\\
\label{newg}
g&=&\frac{a+i\cos\theta}{\sqrt{ b^2+\left(a+i\cos\theta\right)^2 }},
\end{eqnarray}
where $a$ and $b$ depend on $\avg$ and $\avf$:
\begin{eqnarray}\label{Eilenbergera}
a&=&(2\omegan+\avg/\taup)/uv,\\
\label{Eilenbergerb}
b&=&(2\Delta+\avf/\taum)/uv.
\end{eqnarray}
%------------------------------------eq for <f> and <g>
Integrating Eqs. (\ref{newf}) and (\ref{newg}) over $\theta$,
we obtain two self-consistent equations for $\avf=-b~{\rm Im}\sinh^{-1}\big((a+i)/b\big)$ and $\avg=\mbox{Im}\sqrt{b^2+(a+i)^2}$. The equations for $\avf$ and $\avg$ can be recasted in the following form:
\begin{eqnarray}\label{avfeq}
\avf=b\tan^{-1} \frac{\avg}{a},\\
\label{avgeq}
%1&=&\frac{b^2}{1-\avg^2}-\frac{a^2}{\avg^2}.
\avg^4+\avg^2(a^2+b^2-1)-a^2=0.
\end{eqnarray}
%--------------------------------introduciong X and its equation
It is convenient to introduce a new variable $X$ such that
\begin{gather}
X=\frac{\avf uv \taum }{2\Delta\taum+\avf},
\label{X} \\
\avg=\frac{2\omegan\taup}{uv\taup\cot X -1}.
\label{solavg}
\end{gather}
Substituting this into Eqs. (\ref{avfeq}) and (\ref{avgeq}) gives the following equation for $X$
\begin{eqnarray} \label{solX}
\left( \frac{\Delta \sin X \taum}{uv\taum -X}\right)^2+
\left(\frac{\omegan\taup}{uv\taup\cot X-1} \right)^2=\frac14
\end{eqnarray}
The gap equation and the current density $J=|{\bf J}|$ in Eqs. (\ref{gapeq0}) and (\ref{jeq0}) can be expressed in terms of
$X$ as follows:
\begin{gather}
\ln \frac{T}{T_c}+2\pi T \sum_{n=0}^{\infty} \left\{ \frac1{\omegan}
-\frac{2X\taum}{uv\taum-X} \right\}=0,
\label{gapeq} \\
J=4 \pi T e v N(0)\taum^2 \Delta^2  \sum_{n=0}^{\infty}
\frac{ 2X-\sin2X}{\left(uv\taum-X\right)^2}.
\label{jeq}
\end{gather}

Equations (\ref{X})-(\ref{jeq}) implicitly define $\avf$, $\avg$, and
$\Delta$ as functions of $J$ in a superconductor
with arbitrary concentration of nonmagnetic and magnetic impurities in the entire range $0<T<T_c$.
Generally, Eqs. (\ref{X})-(\ref{jeq}) can be solved only numerically, but
analytical solutions can be obtained in some limiting cases.

To make the effect of impurities on $H_s$ more clear, we first discuss the instructive case of
a clean superconductor $(\taup,\taum)\rightarrow\infty$. \cite{Parmenter62,Bardeen62,Makigapless}
In the absence of current $u\rightarrow0$,
Eqs. (\ref{X}) and (\ref{solX}) yield the Gor'kov result, $\avf=\Delta/\sqrt{\Delta^2+\omegan^2}$.
Solving Eqs. (\ref{X}) and (\ref{solX}) for a current-carrying state, $u>0$, gives:
\begin{gather}
\avf=\frac{2\Delta}{uv}\tan^{-1}\sqrt{z/2}
\label{cleanavf} \\
z=\frac{u^2v^2}{4\omegan^2}-\frac{\Delta^2}{\omegan^2}-1+
\left(\left(\frac{u^2v^2}{4\omegan^2}-\frac{\Delta^2}{\omegan^2}-1\right)^2+\frac{u^2v^2}{\omegan^2}\right)^{1/2}
\label{z}
\end{gather}
The dependence of $\Delta$ on $u$ and $T$ is then determined by
Eqs. (\ref{gapeq0}), (\ref{cleanavf}) and (\ref{z}).

Let us discuss the clean limit at $T=0$ in more detail.
For $T\rightarrow0$, the summation in Eqs. (\ref{gapeq}) and (\ref{jeq})
can be replaced by integration. Substituting Eq. (\ref{cleanavf}) in
those equations gives: \cite{Makigapless,Kupriyanov80}
\begin{equation}
\ln\frac{\Delta}{\Delta_{00}}=
\left\{
  \begin{array}{l l}
    	0, & 	\quad \text{$w\leq1$ }\\
		-\cosh^{-1}{w}+\sqrt{1-1/w^{2}},	&	\quad \text{$w>1$}\\
  \end{array}
\right.
\end{equation}
where $w=uv/2\Delta$, $\Delta$ is function of $u$,
and $\Delta_{00}$ is the order parameter at $u=0$ and $T=0$.
Calculating $J(u)$ in Eq. (\ref{jeq}) yields
\begin{equation}
\frac{J}{J_0}=
\left\{
  \begin{array}{l l}
    	1, & 	\quad \text{ $w\leq1$ }\\
		1-(1-1/w^{2})^{3/2},	&	\quad \text{$w>1$}.\\
  \end{array}
\right.
\label{JM}
\end{equation}
Here the current density \begin{equation}\label{J0}\nonumber
J_0=\frac13 eN(0) v^2u = env_s.
\end{equation}
corresponds to superconducting flow of all electrons.

\begin{figure}[ht]
\includegraphics[width=8.5cm,angle=0]{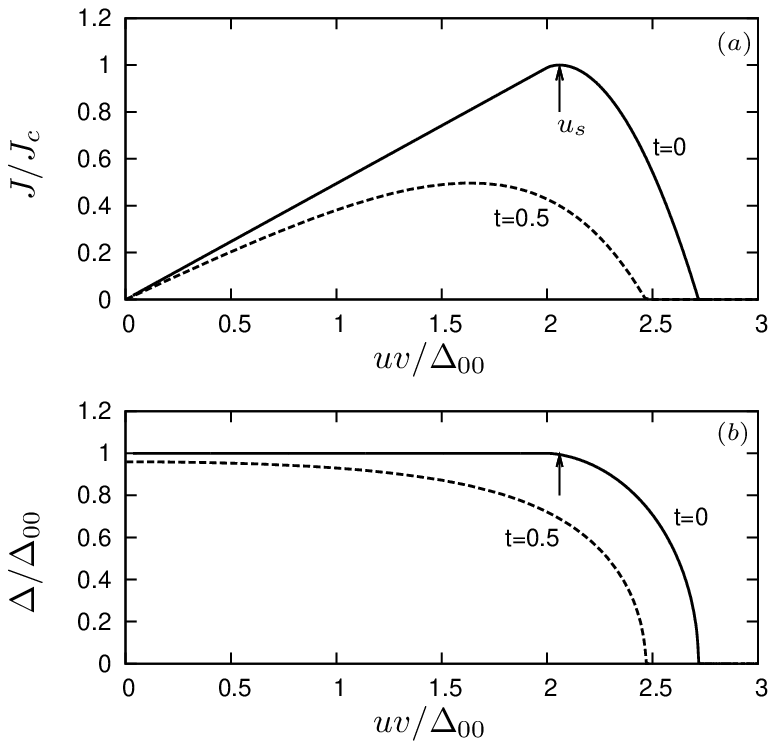}
    \caption{$(a)$ Current density and $(b)$ gap as functions of $u$ at $T\rightarrow0$ (solid line) and at $T=0.5 T_c$ (dashed line).
For $T=0$, $J(u)$ increases linearly with $u$, reaches $J_c$ at $\ush$
and decreases to zero. The gap stays constant for $u\le2\Delta_{00}/v$
and then drops. The arrow shows the location of the deparing momentum $\ush=2.059\Delta_{00}/v$
where current reaches the critical current $J_c$.
At $T=0.5 T_c$, $\ush$ is shifted to lower $u$.}
    \label{fig2}
\end{figure}

The calculated dependencies of $J(u)$ and $\Delta(u)$ on $u$ for a clean superconductor
at $T\rightarrow0$ are shown in \figref{fig2}.
The current $J=env_s$ increases linearly as $u=2m v_s/\hbar$ increases up to $u=2\Delta_{00}/v$, then $J(u)$ becomes nonlinear and reaches
the maximum value $J_c$ at the pair-breaking momentum $\ush=2.059\Delta_{00}/v$, and then drops to zero.
At the same time, the gap $\Delta(u)=\Delta_{00}$ remains unaffected by current at $0<u<2\Delta_{00}/v$, but rapidly
drops to zero for $u>2\Delta_{00}/v$. Here $\Deltash$ at $u=\ush$
is slightly smaller then $\Delta_{00}$.

A clean s-wave superconductor at $T=0$ exhibits no current pairbreaking in the broad range of $0<u<2\Delta_{00}/v$
where the linear London electrodynamics is applicable. For finite temperatures, this anomalous feature no longer holds,
as illustrated by Fig. 2 which shows $J(u,T)$ and $\Delta(u,T)$ calculated numerically from Eqs. (\ref{gapeq})-(\ref{z}) for $T=0.5T_c$.
Here $\Delta(u)$ decreases as $u$ increases, and $J(u)$ deviates from linearity
for $u$ smaller than $\Delta_{00}/v$, similar to the nonlinearity of $J(u)\propto u(1-u^2\xi^2)$ due to current
pairbreaking in the GL theory. Such a difference in the behaviors of $\Delta(u)$ at $T=0$ and $T\sim T_c$ is due to
thermally-activated quasiparticles that are negligible at $T\ll T_c$ but become essential at higher temperatures. In any case, $J(u)$ reaches the
maximum at the critical superfluid momentum $u\sim v/\Delta_{00} \sim 1/\xi_0$, where $\xi_0=\hbar v/\pi \Delta_{00}$ is the superconducting coherence length in the clean limit. The maximum of $J(u)$ at the critical momentum $\ush$ defines the superheating field which will be calculated in the next section.

\begin{figure}[ht]
\includegraphics[width=8.5cm,angle=0]{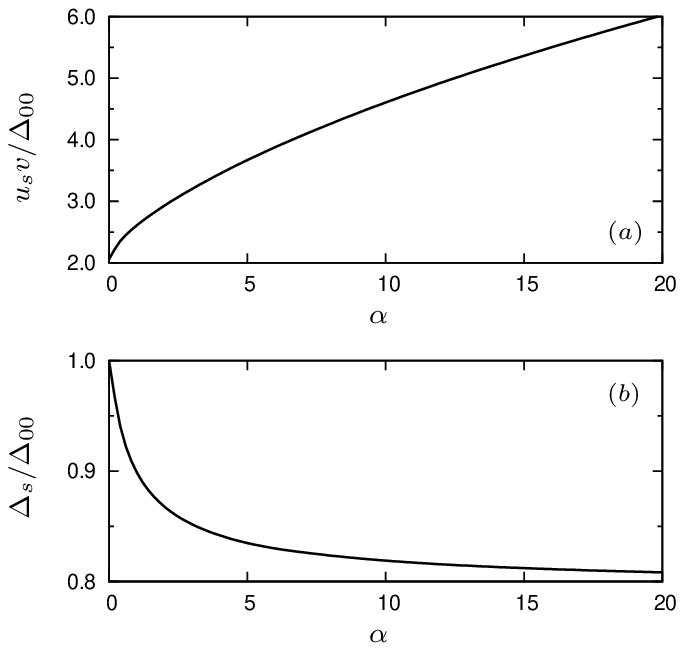}
    \caption{ (a) The pairbreaking momentum $\ush(\alpha)$ as a function of the nonmagnetic scattering rate $\alpha$ at $T=0$.
    (b) $\Deltash$ at the critical momentum $u=u_s$ as a function of $\alpha$.
Here $\Deltash$ approaches $0.79 \Delta_{00}$ for $\alpha\gg 1$.
}
    \label{fig3}
\end{figure}

Impurities do not change qualitatively the behavior of $J(u)$
shown in \figref{fig3}, but they increase the critical momentum $\ush$.
We calculated the effect of nonmagnetic impurities on $\ush$ at
$T\rightarrow0$ by solving the full Eqs. (\ref{solX}), (\ref{gapeq})
and (\ref{jeq}). The results are shown in \figref{fig3} which displays $\ush$
as a function of the scattering parameter $\alpha\equiv1/\Delta_{00} \tau = \pi\xi_0/\ell$
where $\ell = v\tau$ is the mean free path.
The critical momentum $\ush$ increases as $\alpha$ increases,
approaching the square root dependence $\ush\propto\sqrt\alpha$ for $\alpha>1$,
consistent with the general result, $\ush\sim 1/\xi$, where $\xi\simeq (\xi_0\ell)^{1/2}$ in the dirty limit.
The dependence of $\ush(\alpha)$ can be described by the interpolation formula
\begin{equation}
\ush\approx \big(1.053(\alpha+0.655)^{1/2}+1.146\big)\Delta_{00}/v,
\label{ushh}
\end{equation}
which approximates the calculated $\ush(\alpha)$ for $0<\alpha<20$ to an accuracy better than $1\%$.

Numerical solutions of the equation for $\Delta$ at $T=0$ show that in the presence of
impurities $\Delta(u)$ decreases as $u$ increases even if $u\xi\ll 1$.
As a result, nonmagnetic impurities become pairbreakers, reducing the superfluid
density as $J$ increases. \cite{Makigapless} This manifests itself in a
nonlinear Meissner effect even at $T=0$. \cite{Groll}
Shown in \figref{fig3} are the calculations of
$\Deltash$ at the depairing momentum $\ush$ as a function of the
scattering parameter $\alpha$ at $T=0$. Here $\Deltash$ decreases
as $\alpha$ increases, approaching $\Deltash(\alpha\to\infty) = 0.79 \Delta_{00}$
in the dirty limit.

\section{Superheating field}

We calculate $H_s$ for a superconductor occupying the half-space $x>0$ by solving the Maxwell equation
\begin{equation}\label{maxwell0}
-\frac{\partial B}{\partial x}=\frac{4\pi}{c} J,
\end{equation}
where the field is applied along the $z-$ axis, so both the current density $J(u)$ and
the vector potential $A_y$ have only the $y-$ components.
Here $J(u)=-c\delta \mathcal F/\delta A_y$ in Eq. (\ref{jeq0}) can be
obtained by varying the free energy density \cite{Eilenberger68}
%\begin{widetext}
\begin{gather}
\mathcal F = \Delta^2 N(0) \ln\frac{T}{T_c}+2\pi N(0)T \sum_{n=0}^{\infty} \bigg\{
\frac{\Delta^2}{\omegan}-
2\Delta\avf-\nonumber \\
2\omegan(\avg-1)-iuv\avgcos
-\frac12\left(\frac{\avf^2}{\taum}+\frac{\avg^2-1}{\taup} \right)
\!\bigg\},\label{FreeEngorg}
\end{gather}
%\end{widetext}
where $f$ and $g$ are the solution of Eqs. (\ref{EilenbergerEquation0}) and (\ref{normalization}).
For $\kappa\gg 1$, the amplitudes $\Delta$ and $f$ vary slowly on the scale of
$u_s\sim 1/\xi$ so the gradient terms $\nabla f$ in $\mathcal F$ are negligible. In this case
the variational derivative $J=-c\delta \mathcal F/\delta A$ becomes
the partial derivative, $J=-c\partial \mathcal F/\partial A$.
Moreover, since $d\mathcal F/dA=\partial \mathcal F/\partial A + (\partial \mathcal F/\partial\Delta)(\partial \Delta/\partial A) + (\partial \mathcal F/\partial f)(\partial \Delta/\partial f)$ where $\partial \mathcal F/\partial \Delta =\partial \mathcal F/\partial f = 0$ in equilibrium, we have $J=-cd\mathcal F/dA$.

Multiplying both sides of Eq. (\ref{maxwell0}) by $B=dA/dx$ and integrating from $x=0$ to $x=\infty$, we obtain
\begin{equation}\label{maxwell1}
-\int_0^\infty B\frac{dB}{dx} \td x
=-4\pi\int_0^\infty\frac{d \mathcal F}{d A} \frac{d A}{dx} \td x.
\end{equation}
Here the boundary conditions are: $B\rightarrow 0$
and $\mathcal F \rightarrow -H_c^2/8\pi$ at $x\rightarrow \infty$, and
$B\rightarrow H$ at $x=0$. Thus, Eq. (\ref{maxwell1}) reduces to:
\begin{equation}\label{sh}
H^2=H_c^2+8\pi\mathcal F[u(0)],
\end{equation}
where $H_c$ is the thermodynamic critical field and
$\mathcal F[u(0)]$ and $u(0)$ are the free energy density and the momentum of condensate at the surface, respectively. Eq. (\ref{sh})
thus determines the dependence of $u(0)$ on the applied field.

As the magnetic field $H$ increases, $J(0)$ and $u(0)$ at the surface increase. However, as $u(0)$ reaches $u_s$ at which $J(u)$ is maximum,
further increase of $H$ does not cause any increase of screening current density, making the Meissner state absolutely unstable with respect to small perturbations of the order parameter and initiating penetration of vortices. The condition $u(0)=\ush$ in Eq. (\ref{sh}) defines the superheating field:
\begin{equation}\label{shf}
\Hsh^2(T)=H_c^2(T)+8\pi\mathcal F (\ush,T).
\end{equation}
Substituting the gap equation (\ref{gapeq0})
and $2i\avgcos=a \big(\avg+1/\avg\big)-b\avf$ to Eq. (\ref{FreeEngorg}) yields
$\mathcal F = 2\pi N(0)T \sum_{\omegan>0} \omegan
\big(2-\avg-1/\avg\big)$. Hence, we obtain the final expression for $H_s$:
\begin{equation}\label{shff}
\Hsh^2=H_c^2+16\pi^2 N(0)T\sum_{n=0}^{\infty} \omegan
\left\{2-
\langle g_s\rangle-\frac{1}{\langle g_s\rangle}\right\}.
\end{equation}
Here the index $s$ means that the function $\avg$ is calculated at the critical momentum $u=u_s$.
The thermodynamic critical field $H_c$ is given by
\begin{equation}\label{hcstatic}
H_c^2=16 \pi^2 N(0) T \sum_{n=0}^{\infty}
\left\{
\frac{2\omegan^2+\Delta_{0}^2}{\sqrt{\omegan^2+\Delta_{0}^2}}-2\omegan
\right\}
\end{equation}
where $\Delta_0(T)$ is the superconducting order parameter at $u=0$.
In the clean limit, Eq. (\ref{shff}) reduces to:
\begin{gather}\label{CleanHsh}
\Hsh^2 =
16\pi^2 T N(0)\sum_{n=0}^{\infty}
\bigg\{
\frac{2\omegan^2+\Delta_0^2}{\sqrt{\omegan^2+\Delta^2_{0}}} %\nonumber \\
-\frac{\ush^2 v^2+2\omegan^2\Ysh}{\ush v\sqrt{2\Ysh}}
\bigg\}
\end{gather}
where $z(u)$ is given by Eq. (\ref{cleanavf}). \cite{Galaiko66}

Equations (\ref{shff}) and (\ref{hcstatic}) combined with Eqs. (\ref{solX})-(\ref{jeq}) define
$H_s$ as a function of temperature, and concentrations of nonmagnetic and magnetic impurities. Calculations of $H_s$ involves solving coupled Eqs. (\ref{solX}), (\ref{gapeq}), (\ref{jeq}) to obtain $\Delta$ and $J$ as functions of $u$ and then finding self-consistently the depairting momentum $u_s$ from the condition $dJ/du=0$. The results of these calculations of $H_s$ for different concentrations of impurities and temperatures are presented below.

\subsection{Effect of nonmagnetic impurities}
%---------------------------------nonmagetic impurities---------------------

We quantify the effect of nonmagnetic impurities on $H_s$ by the dimensionless scattering rate
\begin{equation}
\alpha = \hbar/\tau\Delta_{00}=\pi\xi_0/\ell,
\label{alp}
\end{equation}
where $\xi_0=\hbar v/\pi\Delta_{00}$ is the clean limit coherence length, and $\ell = v\tau$ is the mean free path.
For $u=0$, impurities do not affect $\Delta$, $H_c$ and $T_c$ of s-wave superconductors \cite{Anderson59,Abrikosov61}, but
in a current state both $\Delta$ and $u_s$ are affected by impurities \cite{Makigapless}, as shown in \figref{fig3}.

\begin{figure}[ht]
\includegraphics[width=8.5cm,angle=0]{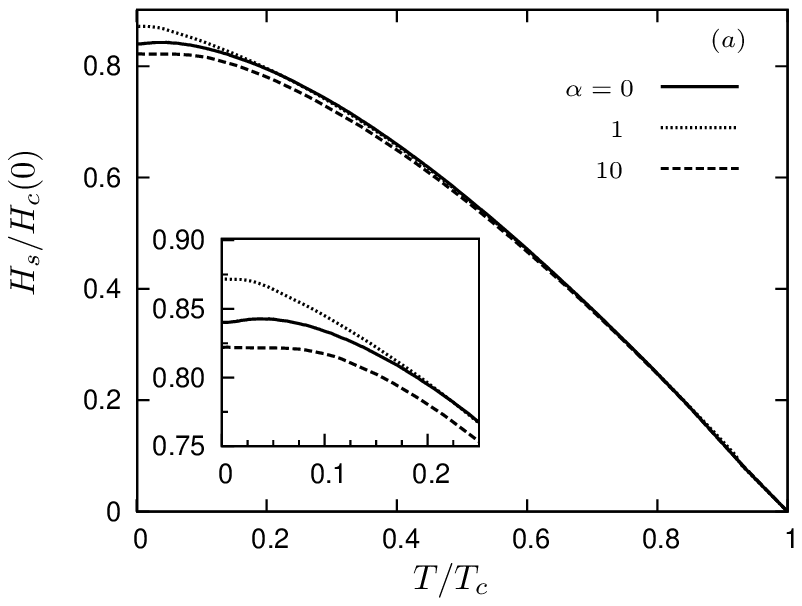}
\includegraphics[width=8.5cm,angle=0]{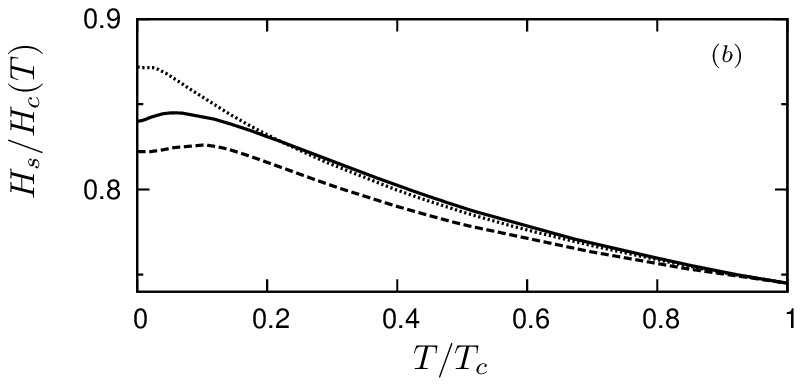}
\caption{Temperature dependencies of $\Hsh$ calculated for $\alpha=0,1$ and $10$.
(a) $\Hsh(T)$ in the units of $H_c(0)$ and (b) the ratio $\Hsh(T)/H_c(T)$ where $H_c(T)$ is calculated from Eq.(\ref{hcstatic}).
At low $T$, nonmagnetic impurities result in nonmonotonic dependence of $H_s$ on $\alpha$, eliminating the maximum in $\Hsh(T)$ at
$T=0.04T_c$ for the clean limit, as shown in the inset. The difference between $\Hsh(T)$ diminishes at higher $T$ where the $\Hsh(T)$ curves approach the GL result given by Eq. (\ref{h1}).
}
    \label{fig4}
\end{figure}

We first consider the effect of nonmagnetic impurities on the temperature dependence of $\Hsh(T)$. Shown in \figref{fig4} are the superheating fields calculated for the clean ($\alpha=0$), moderately dirty ($\alpha=1$) and dirty ($\alpha=10$) limit. The first conclusion apparent from \figref{fig4} is that nonmagnetic impurities have a rather weak effect on $\Hsh$, despite their pairbreaking nature in the presence of current. Near $T_c$ the $\Hsh(T)$ curves coincide, approaching $\Hsh(T)=(\sqrt{5}/3)H_c$ predicted by the GL theory, where $H_c$ is independent of $\alpha$ according to the Anderson theorem \cite{Anderson59}. At lower temperatures impurities result in different behaviors of $\Hsh(T)$ for different $\alpha$. For instance, one of the features of the clean limit at low temperatures is a maximum in $\Hsh(T)$ at $T=0.04 \TcCleanStatics$ pointed out by Catelani and Sethna.\cite{Catelani08} As follows from \figref{fig4}, this maximum disappears in the moderately dirty limit ($\alpha = 1$), as well as the dirty limit ($\alpha=10$). As will be shown below, this change in the behaviors of $\Hsh(T)$ at low temperatures results from the impurity-induced change in the quasiparticle density of states at $H=H_s$ from the gapless state in the clean limit to a gapped state for $\alpha >\alpha_c$.

\begin{figure}[ht]
\includegraphics[width=8.5cm,angle=0]{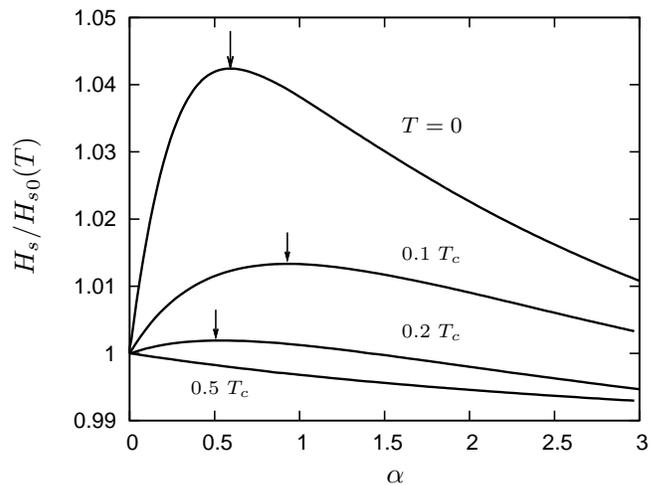}
    \caption{Enhancement of $\Hsh$ by nonmagnetic impurities at low temperature where
$\Hsh(\alpha,T)$ is normalized to ${\Hsh}_0(T)$ for a clean superconductor.
Here the position of the maximum $\alpha_{max}(T)$ in $\Hsh(\alpha)$ depends on temperature: $\alpha_{max}=0.59$ for $T=0$, $\alpha_{max}=0.91$ for $T=0.1T_c$, and
$\alpha_{max}=0.53$ for $T=0.2T_c$. At higher temperatures, the maximum disappears.}
    \label{fig5}
\end{figure}

We now turn to the dependence of $\Hsh$ on the nonmagnetic scattering rate $\alpha$.
Shown in \figref{fig5} are $\Hsh(\alpha)$ calculated for different temperatures. These $H_s(\alpha)$ curves exhibit surprising
maxima at low temperatures, resulting in a $\simeq 4.2\%$ enhancement of $\Hsh(\alpha)$ at $T=0$ as compared to the clean limit.
Here the arrows in the figure mark the maxima in $\Hsh$, the location of which depends nonmonotonically on temperature as described in the caption to
\figref{fig5}. Thus, there is an optimum mean free path $\ell_{max}=5.3\xi_0$ at $T=0$ for which the superheating field is maximum. At higher temperatures, $T\gtrsim 0.3T_c$, the maximum in $\Hsh(\alpha)$ disappears.

%%%%%%%%%%%%%% Fig_Hsh vs nonmagnetic impurities %%%%%%%%%%%%%%

\begin{figure}[ht]
\includegraphics[width=8.5cm,angle=0]{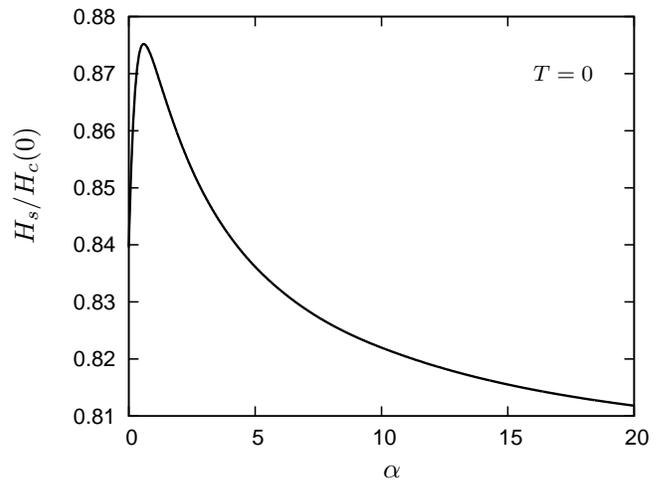}
    \caption{Dependence of $H_s$ at $T=0$ on the nonmagnetic scattering rate $\alpha$.}
    \label{fig6}
\end{figure}

The results of this section show that the effect of nonmagnetic impurities on $\Hsh$ is most pronounced at low temperatures where impurities can eliminate the maximum in the temperature dependence of $\Hsh$ characteristic of the clean limit. At the same time, impurities can cause a new maximum in $\Hsh$ as a function of the scattering rate $\alpha$ at low temperatures, although the overall effect of nonmagnetic scattering on $\Hsh$ turns out to be comparatively weak. This  results from two opposite effects that nearly cancel each other: the increase of $\Hsh$ due to the increase of the pairbreaking momentum $u_s$ shown in \figref{fig3} and the decrease of superfluid density as $\alpha$ increases. Here $\Hsh$ roughly scales like the thermodynamic critical field $H_c$ in which these opposite trends cancel out exactly, as prescribed by the Anderson theorem. Yet our results show that this cancelation is not exact for $\Hsh$ which depends weakly on the scattering rate $\alpha$, as illustrated by \figref{fig6} which shows the calculated $\Hsh(0,\alpha)$ at $T=0$. For instance $H_s \approx 0.82 H_c$ at $\alpha=10$ is rather close to $H_s=0.84H_c$ in the clean limit.

%-----------------------------------------Discussing magnetic impurities
\subsection{Effect of magnetic impurities}
Unlike nonmagnetic impurities, magnetic impurities strongly suppress $T_c$, \cite{Abrikosov61}
resulting in a significant reduction of $\Hsh$ as well. We quantify this effect by
the dimensionless magnetic scattering rate $\alpha_m$ similar to that was used above for nonmagnetic impurities:
\begin{equation}
\alpha_m=\hbar/\tau_m\Delta_{00} = \pi\xi_0/\ell_m
\label{alm}
\end{equation}
where $\tau_m$ and $\ell_m=v\tau_m$ are the spin-flip scattering time and mean free path, respectively.

\begin{figure}[ht]
\includegraphics[width=8.5cm,angle=0]{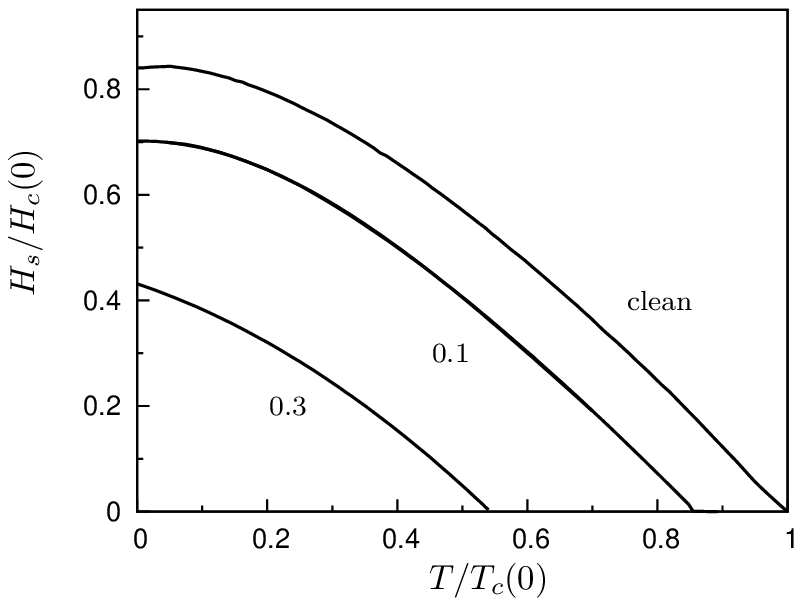}
    \caption{Temperature dependencies of $\Hsh(T)$ calculated for $\alpha=0$ and different magnetic scattering rates $\alpha_m$. Here $T_c(0)$ is the critical temperature for clean superconductors.
	}
    \label{fig7}
\end{figure}

Shown in \figref{fig7} are the temperature dependencies $\Hsh(T)$ calculated for $\alpha=0$ and different values
of $\alpha_m$. The superheating field is significantly reduced by magnetic scattering, the $\Hsh$ suppression is stronger at higher
temperatures. One can see that for $\alpha_m\simeq 0.1$, which corresponds to $\ell_m\sim 30\xi_0$, the superheating field is roughly
reduced by half as compared to the case of $\alpha_m=0$.

\begin{figure}[ht]
\includegraphics[width=8.5cm,angle=0]{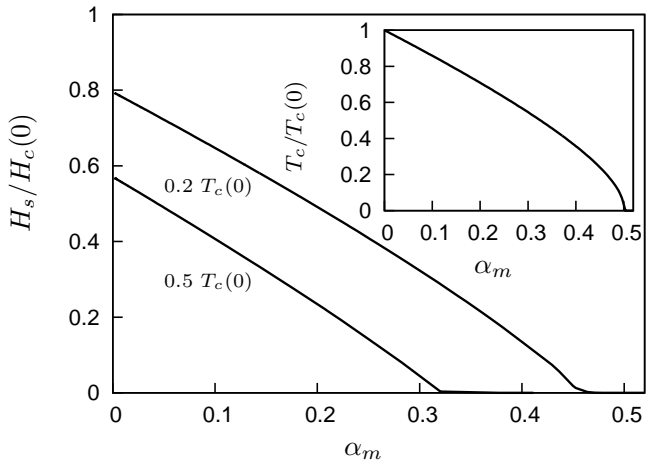}
    \caption{Dependencies of $\Hsh$ on $\alpha_m$ calculated for at $T=0.2T_c(0)$ and $T=0.5T_c(0)$
	Insert shows the dependence of $T_c$ on $\alpha_m$ calculated from the Abrikosov-Gor'kov theory \cite{Abrikosov61}
    }
    \label{fig8}
\end{figure}

Figure \ref{fig8} shows the reduction $\Hsh(\alpha_m)$ by magnetic impurities at a given
temperature. Unlike the effect of nonmagnetic impurities which is most pronounced at low temperatures, the effect of magnetic
impurities is stronger at higher temperature. Here $\Hsh(\alpha_m)$
decreases as $\alpha_m$ decreases, vanishing at a critical scattering rate $\alpha_{mc}(T)$.

\begin{figure}[ht]
\includegraphics[width=8.5cm,angle=0]{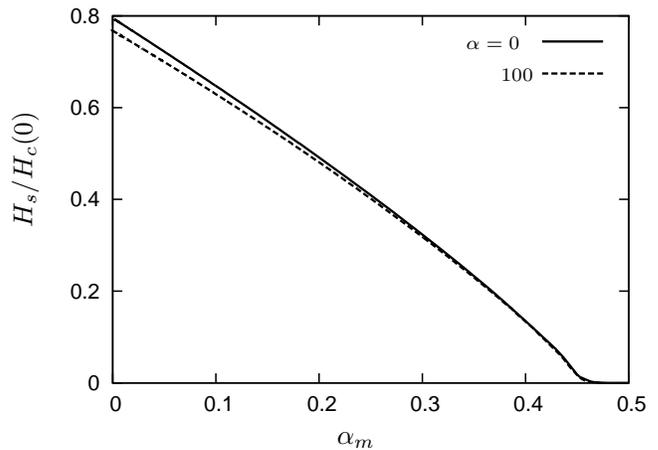}
    \caption{$\Hsh(\alpha_m)$ calculated for $T=0.2T_{c}(0)$ and different nonmagnetic scattering rates $\alpha$.
	}
    \label{fig9}
\end{figure}

The combined effect of magnetic and nonmagnetic impurities on $\Hsh$ is
shown in \figref{fig9}. From the curves $\Hsh(\malpha)$ calculated for different $\alpha$ at $T=0.2T_c(0)$,
we can see that there is practically no interplay between the effects of magnetic and nonmagnetic scattering
on $\Hsh$. Similar to the case of $\alpha_m=0$, the effect of nonmagnetic impurities at $\alpha_m>0$ is most noticeable at low temperatures,
where it remains rather weak even for very dirty superconductors with $\alpha \sim 100$.

\section{Quasiparticle Density of States in the Meissner State}\label{DoS}

As was discussed in the Introduction, the effect of current on the
quasiparticle density of states $N(\epsilon)$ in the Meissner state
is rather nontrivial because a clean superconductor at $H=\Hsh$ is in a gapless state.
The appearance of quasiparticle at the Fermi level at $\Hsh$ can have important
consequences for the breakdown of superconductivity under strong
RF fields, as will be discussed below. This feature of the s-wave
clean limit results from the spectrum of quasi-particles in a uniform
current-carrying state \cite{Parmenter62,Bardeen62}:
\begin{equation}\label{spectrum}
\varepsilon_{\bf k}=\sqrt{\Delta_0^2+v^2(p-p_F)^2}+\hbar{\bf u}\cdot{\bf v}/2
\end{equation}
where the last term describes the effect of current with the superfluid velocity $v_s=\hbar u/2m$.
The quasiparticle gap $\epsilon_g$ corresponds to the minimum of $\varepsilon_{\bf k}$, giving
$\epsilon_g=\Delta_0$ for $u=0$. However, in the presence of current, $\Delta_0(T)$ is no longer
the gap in the quasiparticle spectrum. For instance, $\Delta_{00}$ at $T=0$
remains independent of $J$ in the entire interval $0<u<2\Delta_{00}/\hbar v$,
as shown in \figref{fig2}, while the anisotropic quasi-particle gap $\varepsilon_g({\bf u}) = |\Delta_{00}| + \frac{1}{2}\hbar uv\cos\varphi$, obtained by setting $p=p_F$ in Eq. (\ref{spectrum}), depends on the angle $\varphi$ between the current and quasiparticle momentum. The minimum gap $\varepsilon_g=|\Delta_{00}|-\hbar uv/2$ vanishes at the condensate momentum $u>u_g=2\Delta_{00}/\hbar v$ slightly smaller than the pairbreaking momentum $u_s$ corresponding to $H=\Hsh$. The question is then how the density of states at $H=\Hsh$ is altered by impurity scattering. To address this issue we use the approach developed by Maki \cite{Maki63,Makigapless} and Fulde \cite{Fulde65} to calculate $N(\epsilon,u)$ as functions of energy $\epsilon$ and the superfluid velocity. For the sake of simplicity, we only consider here the effect of nonmagnetic impurities on $N(\epsilon,u)$.

The density of state $N(\epsilon)$ is given by the imaginary part
of the real frequency Eilenberger function $g(\epsilon)$ obtained
by analytic continuation of the thermodynamic function $g(\omega_n)$
from the imaginary Matsubara axis onto the the real energy axis
$\epsilon$. \cite{Kopnin}
We consider here the
limit $T=0$ for which this procedure yields \cite{Maki63,Fulde65}:
\begin{eqnarray}\label{DoSMaki}
\DoS(\epsilon)&=&{\rm Im} \frac{2uv\tau^2\epsilon}{uv\tau- \tan \chi},
\end{eqnarray}
where $\nu(\epsilon)=N(\epsilon)/N(0)$ is the normalized density of states, and $\chi$ is the solution of the real-frequency version of Eq. (\ref{solX}) in which the substitution $\omegan\to -i\epsilon$ is made:
\begin{equation}
\label{DosEqn}
%\Delta^2\left(\frac{uv}{\sin\chi}-\frac1{\tau} \frac{\chi}{\sin\chi}\right)^{-2}
%-\romega^2\left(\frac{uv}{\tan\chi}-\frac1{\tau}\right)^{-2}=\frac14.
\left( \frac{\Delta \sin \chi \tau}{uv\tau -\chi}\right)^2-
\left(\frac{\romega\tau}{uv\tau\cot \chi-1} \right)^2=\frac14
\end{equation}
Eqs. (\ref{DoSMaki}) and (\ref{DosEqn}) allow us to obtain $N(\epsilon)$ using $\Delta(u)$ and $u_s$ calculated in the previous sections with the use of the thermodynamic Eilenberger equations.

\begin{figure}[ht]
        \includegraphics[width=8.5cm,angle=0]{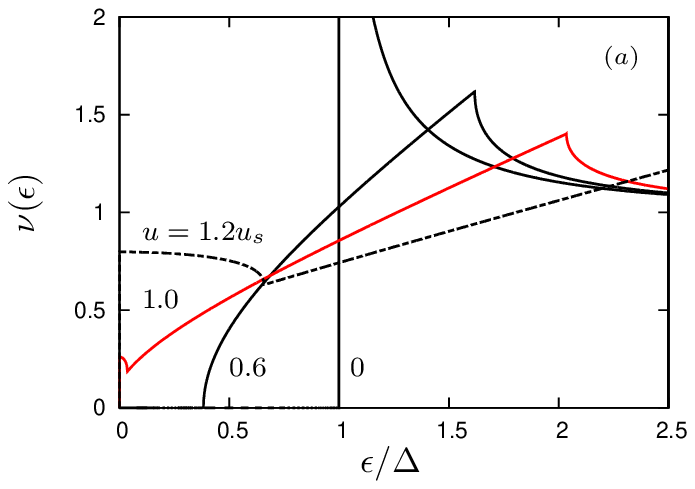}
 	 \includegraphics[width=8.5cm,angle=0]{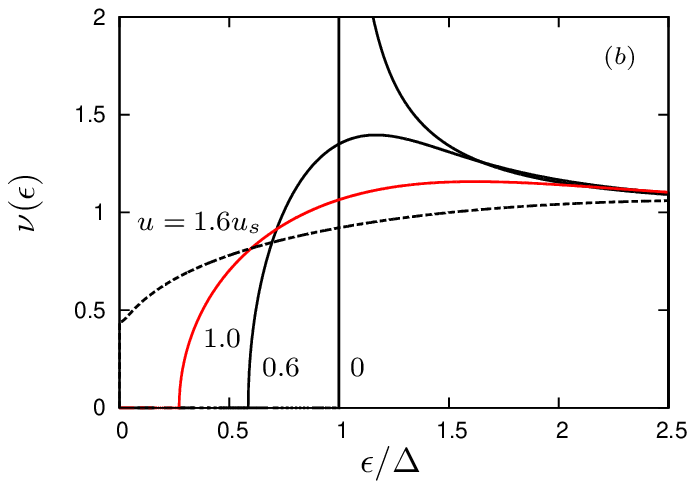}
        \caption{Density of state $\nu(\epsilon)$ in the current-carrying states for:
		     	  	(a) clean limit, $\alpha=0$;
        			(b) moderate dirty limit, $\alpha=3.6$.
                Here $\Delta$ is understood as $\Delta(u)$ at $T=0$ for a given $\alpha$.
				The red lines show $\nu(\epsilon)$ at $u=\ush$.
				Solid lines show $\nu(\epsilon)$ for $u<\ush$,
				while dashed lines correspond to $u>\ush$.
				For $\alpha=0$, the gap in the spectrum closed
				at $u>0.970\ush$. For $\alpha=3.6$, we obtained $\gapsh=0.211\Delta$.
				}
        \label{fig10}
        \end{figure}

Figure \ref{fig10} $(a)$ shows the evolution of $\nu(\epsilon)$ calculated for a clean superconductor $(\alpha=0)$ as the condensate wave vector $u$ increases. For $u=0$, we recover the BCS density of states $\nu(\epsilon)=\epsilon/\sqrt{\epsilon^2-\Delta_0^2}$ for $\epsilon>\Delta_0$ and
$\nu(\epsilon)=0$ for $\epsilon<\Delta_0$. In the presence of current, the singularity in $\nu(\epsilon)$ at $\epsilon=\Delta_0$ disappears, and
the energy gap in the spectrum $\epsilon_g(\alpha,u)$ defined as the maximum energy at which $\nu(\epsilon_g)$ vanishes, becomes smaller than $\Delta_0$. Here the quasiparticle gap $\epsilon_g(u)$ decreases as $u$ increases. At the critical value $u=\ush$ corresponding to the superheating field, the density of state at the Fermi surface in the gapless state equals $\DoS(0)=0.243$.

The effect of impurities on $\nu(\epsilon, u)$ for a moderately dirty
superconductor with $\alpha=3.6$ is shown in \figref{fig10} $(b)$. For $u=0$, nonmagnetic impurities do not change the BCS density of states,
in accordance with the Anderson theorem. However, $\nu(\epsilon)$ in a current-carrying superconductor with impurities begins to differ markedly from $\nu(\epsilon)$ in a clean superconductor. As it is evident from \figref{fig10} $(a)$ and $(b)$, impurities not only smear the cusps
in $\nu(\epsilon)$  characteristic of the clean limit but also reduce $N(0)$ at $u=\ush$, eventually restoring the gapped state at
$H=\Hsh$ where $\nu(\epsilon) = 0$ for $\epsilon<\epsilon_g(\alpha)$. For the particular case shown in \figref{fig10} $(b)$,
our calculations give $\gapsh(3.6)=0.211\Delta(u_s)=0.169\Delta_{00}$. For $\alpha\gg 1$, the gap approaches the limiting value $\gapsh=0.410\Delta(u_s)=0.323\Delta_{00}.$ \cite{Fulde65}

%%%%%%%%%%%%%%%%%% DoS for  α=0 and α=1 %%%%%%%%%%%%%%%%%%

\begin{figure}[ht]
	 \includegraphics[width=8.5cm,angle=0]{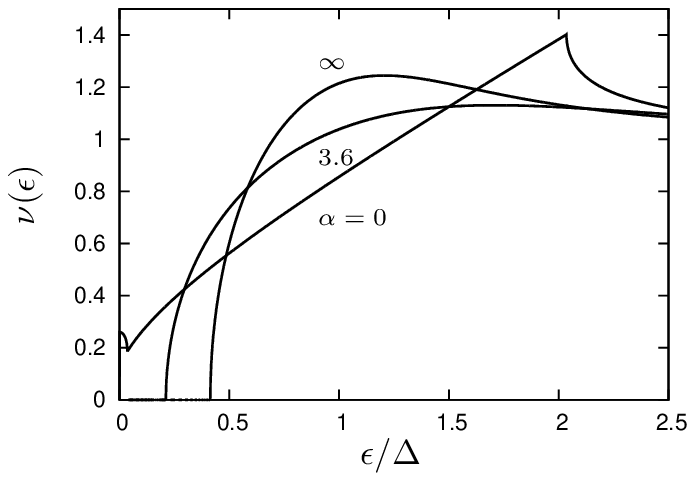}
       \caption{Density of state $\nu(\epsilon)$ at $u=u_s$ for different scattering rates $\alpha$.
		Here $\Delta=\Delta(\alpha)$ at $T=0$ and $u=u_s$.
		}
       \label{fig11}
        \end{figure}

\begin{figure}[ht]
%%%%%%%%%%%%%%%%%%%%%%%%%%%% Gap vs α %%%%%%%%%%%%%%%%%%%%%%%%%%%%
\includegraphics[width=8.5cm,angle=0]{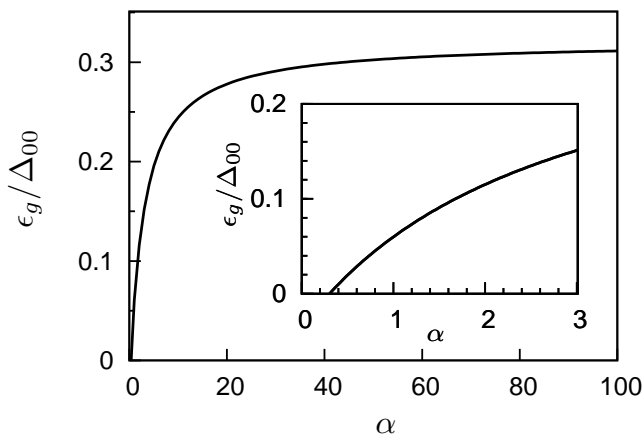}
    \caption{Gap in the quasiparticle spectrum at the depairing momemtum $u=\ush$ as a function of the nonmagnetic scattering rate $\alpha$.
    The gap $\epsilon_s$ opens at $\alpha\ge0.36$.
	(Insert) The behavior of $\epsilon_s(\alpha)$ at small $\alpha$.}
    \label{fig12}
\end{figure}

%-------------------------discussing DoS at Hsh
The different behaviors of $\nu(\epsilon)$ in the clean and dirty limits shown in \figref{fig10}
suggest that at the superheating field a quasiparticle gap $\epsilon_s(\alpha)$
appears as a superconductor gets dirtier. This is illustrated by \figref{fig11} which shows the evolution of $\nu(\epsilon, u_s)$
at the depairing momentum as $\alpha$ increases. One can clearly see the transition from a gapless to a gapped state induced by
nonmagnetic impurity scattering, giving, for example, $\gapsh=0.17\Delta_{00}$ at $\alpha=1$.  Therefore, at $u=u_s$, a quasiparticle
gap $\epsilon_s(\alpha)$ opens at $\alpha>\alpha_c$, where the critical scattering rate $\alpha_c$ calculated numerically from
Eqs. (\ref{DoSMaki}) and (\ref{DosEqn}) is:
\begin{equation}
\alpha_c=0.36
\label{ac}
\end{equation}
The calculated quasiparticle gap $\epsilon_s(\alpha)$ at $u=u_s$
is shown in \figref{fig12}. Here $\epsilon_s(\alpha)$ monotonically increases as
$\alpha$ increases above $\alpha>\alpha_c$, approaching $\gapsh(\infty)\approx 0.323\Delta_{00}$ at $\alpha\to\infty$. The dependence of
$\epsilon_s$ on $\alpha$ can be approximated by the formula
\begin{equation}
\epsilon_g = 0.566\Delta_{00}\big(\tan^{-1}(0.626\alpha + 1.345) -1\big)
\label{egg}
\end{equation}
to the accuracy better than $1.2\%$.


\section{Nonlinear screening and $\Hsh$ in multilayers}
In this section we use the results obtained above to calculate screening of a magnetic field parallel to a multilayer consisting of alternating superconducting (S) and insulating (I) layers, as shown in \figref{fig1}. Such multilayers stabilize the Meissner state against penetration of vortices up to the superheating field of the material of the thin S-layers with $d\ll\lambda$ for which the parallel lower critical magnetic field $H_{c1}=(2\phi_0/\pi d^2)\ln(d/\xi)$ is greatly increased. \cite{Gurevich06apl} In turn, the suppression of perpendicular currents in thin S-layers by non-conducting I-layers also suppresses the pairbreaking instability at the finite wave vectors $k$ along the surface, \cite{Chapman95} which initiates penetration of vortices. As a result, the superheating field $\Hsh$ in thin film multilayers is defined by the condition that $J(u)$ in the first layer reaches the depairing current density.

Penetration of magnetic field is described by the Maxwell equation $\nabla\times\nabla\times {\bf A}=4\pi{\bf J}/c$, where the supercurrent $J(u)$ depends on the gauge invariant phase gradient ${\bf u}=\nabla\varphi+2\pi {\bf A}/\phi_0$ as shown in \figref{fig2} $(a)$. Here $\varphi({\bf r})$ is the phase of the order parameter. In the Meissner state ${\bf B}=\nabla\times{\bf A}= (\phi_0/2\pi)\nabla\times{\bf u}$, the Maxwell equation for the planar multilayer geometry in which ${\bf u}=\big(0,u(x),0\big)$ can be written in terms of the $y$-component $u(x)$:
\begin{equation}
\frac{\partial^2 u}{\partial x^2}=\frac{8\pi^2}{\phi_0c} J(u).
\label{ulay}
\end{equation}
We assume specular scattering of quasiparticles at the S-I interfaces, so there is no suppression of the order parameter due to surface scattering in the S-layers. In this case $J(u)$ is nearly uniform across each S-layer,
a slight decreases of $u(x)$ as $x$ increases resulting from the London screening over the penetration length $\lambda\gg d$. The solution of Eq. (\ref{ulay}) gives the distribution of magnetic field $B(x)=(\phi_0/2\pi)\partial_xu(x)$ across the multilayer.

\begin{figure}[ht]\includegraphics[width=8.5cm,angle=0]{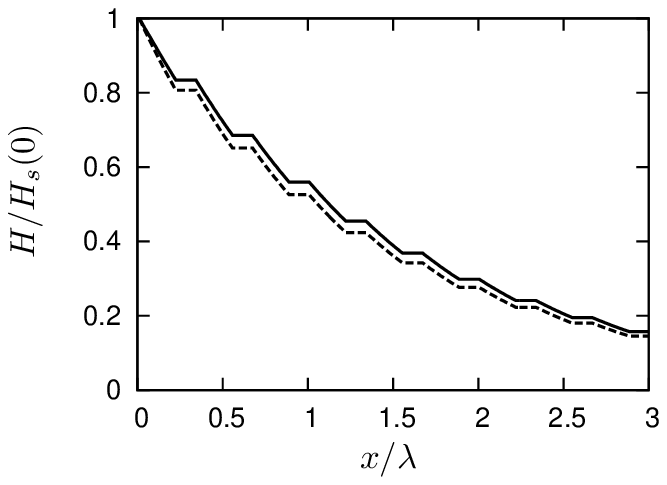}
    \caption{Distribution of the magnetic field $H(x)$ in a multilayer obtained by numerical solution of Eq. (\ref{ulay}) for $\alpha=1$ and $T=0.5T_c$  (solid line). The result of the London model for the same parameters is shown by the dashed lines where
    $\lambda$ is the penetration depth for the material of the S layer.
    }
    \label{fig13}
\end{figure}

The boundary conditions to Eq. (\ref{ulay}) are as follows: $\partial_xu(0)=2\pi H/\phi_0$ at $x=0$, $u(\infty)=0$ at $x\to\infty$, constant $B$ in the I-layers and the continuity of $u(x)$ at the S-I interfaces. Shown in \figref{fig13} is the distribution of $H(x)$ calculated from Eq. (\ref{ulay}) in which $J(u)$ given by Eq. (\ref{jeq})
was calculated using the solution of the Eilenberger equations for $\alpha=1$ and $T=0.5T_c$. As a comparison we also show $H(x)$ (dashed line) calculated by solving the linear London equation $\lambda^2\partial_{xx}u - u=0$ for the same parameters.
One can see that the Eilenberger theory which takes current pairbreaking into account, gives slower penetration of the magnetic field in the first few layers at the surface as compared to the London model which disregards current pairbreaking effects.

A curious manifestation of pairbreaking effects in the Meissner magnetization of moderately clean superconductors occurs due to the transition from the gapped to the gapless state as the field $H$ increases above $H_g$ defined by Eq. (\ref{hg}). Consider for example the magnetization $m(H)$ of a long cylinder of radius $R\gg\lambda$ in a parallel magnetic field:
\begin{equation}
m=\frac{1}{c}\int_0^RJ(v_s)dr
\label{m}
\end{equation}
where we use the dependence of $J(v_s)$ on the superfluid velocity $v_s=v_cH(x)/H_g$ for $T=0$ in the clean limit given by Eq. (\ref{JM}). For $H<H_g$, the dc magnetization $m_0(H)=-H/4\pi$ corresponds to the linear Meissner effect in the entire field region $0<H<H_g$. However, as $H$ exceeds $H_g$ the magnetization exhibits singularities in higher order derivatives indicating a phase transition from the gapped to the gapless state. Indeed, let $H$ be slightly above $H_g$ so that $m(H)=m_0(H)+m_a(H)$ where $m_0(H)=-H/4\pi$ is the ideal Meissner magnetization, and $m_a(H)$ is the nonlinear contribution due to the second term in Eq. (\ref{JM}):
\begin{equation}
m_a=\frac{J_c}{c}\int_0^L(w^2-1)^{3/2}\frac{dx}{w^3}
\label{ma}
\end{equation}
where $w(x)=H(x)/H_g$, $H_g=4\pi J_c\lambda/c$, $H(x)\approx H-Hx/\lambda$ is the field profile at the surface $x=0$, and $L=(H-H_g)\lambda/H_g \ll \lambda$ is the depth of the gapless layer. Equation (\ref{ma}) at $\varepsilon = (H-H_g)/H_g \ll 1$ then  yields $m_a=(\sqrt{2}/5\pi)H_g\varepsilon^{5/2}$, which means a square-root singularity in the third derivative of $m_a(H)$ at $H=H_g+0$:
\begin{equation}
\frac{d^3m_a}{dH^3}=\frac{3}{4\pi\sqrt{2} H_g^2}\left(\frac{H}{H_g}-1\right)^{-1/2}
\label{dm}
\end{equation}
The singularity in the thermodynamic quantity $d^3m/dH^3$ implies the singular discontinuity in the forth derivative of the thermodynamic potential, indicating a field-induced forth order phase transition. Although Eq. (\ref{dm}) was obtained for the clean limit $\alpha = 0$, this singularity remains in the moderately clean limit as well if the scattering rate $\alpha<0.36$ is smaller than the critical value given by Eq. (\ref{ac}) so the transition from the gapped to the gapless state occurs below the superheating field $H=H_g<H_s$. Experimentally this rather weak singularity may be smeared out by local inhomogeneities of impurity concentration that result in a distribution of the local fields $H_g({\bf r})$.

\section{Discussion}

In this paper, we use the Eilenberger equations to calculate the effect of nonmagnetic and magnetic impurities
on the superheating field of type-II superconductors. Unlike magnetic impurities which strongly suppress $\Hsh$,
nonmagnetic impurities affect $H_s$ weakly, although they can cause a nonmonotonic dependence of $H_s$ on the
scattering rate $\alpha$ at low temperatures. For instance, at $T=0$, nonmagnetic impurities can increase
$H_s$ by $\simeq 4\%$ at $\alpha\approx 0.6$ as compared to the clean limit. As the scattering rate $\alpha$ further increases, $H_s(\alpha)$ decreases and levels off at $H_s(\infty)\approx \Hsh(0)$, as shown in \figref{fig6}.
This decrease of $H_s$ at $\alpha\gg 1$ is consistent with the decrease
of the gap parameter $\Delta_s$ at the pairbreaking momentum $u=u_s$ shown in \figref{fig3}. Yet, unlike the decrease of the
depairing current density $J_c\propto \alpha^{-1/2}$ in the dirty limit,\cite{Kupriyanov80} our results show that the superheating field
roughly scales like $H_c$ even at low temperatures, so $H_s$ is weakly affected by nonmagnetic impurity scattering.

 Our results obtained in the limit of $\kappa\gg 1$ give the lower bound of $H_s(T)$. The effect of finite $\kappa$ increases $H_s(T)$ since the condition $v_s(0)=v_ c$ is no longer sufficient to cause the instability of the Meissner state in the surface layer of thickness $\xi$ where the superfluid velocities $v_s(x) \simeq (1-x/\lambda)v_s(0)$ decreases below $v_c$ because of the London screening. The GL calculations \cite{Fink69,Chapman95,Dolgert96,corn} have shown that the finite-$\kappa$ effects increase $H_s(\kappa)\simeq \big(1+ 0.7\kappa^{-1/2}\big)H_s(\infty)$, giving a correction $\simeq 7-16\%$ as compared to $H_s(\infty)$ at $\kappa\to\infty$ for $\kappa = 20-100$. These effects also make $H_s(\kappa)$ dependent on $\alpha$
 since $\kappa(\alpha)$ increases as $\alpha$ increases approaching $\kappa\sim \alpha\kappa_0$ in the dirty limit. 
 Thus, $H_s(T)$ generally decreases as the scattering rate $\alpha$ increases although this effect is comparatively weak for high-$\kappa$ materials. Addressing the dependence of $H_s$ on $\alpha$ due to the finite-$\kappa$ effects at low temperatures requires the instability analysis of the Meissner state with respect to 2D perturbations of $\delta f(x,y)$ and $\delta{\bf J}(x,y)$ described by the linearized Eilenberger equations.

The nonmonotonic dependence of $H_s$ on $\alpha$ obtained in this work results from
interplay of Meissner currents and impurity scattering, and their effect on the quasiparticle density
of states $N(\epsilon,J)$. Our calculations revealed the disorder-induced transition
from the gapless to the gapped state at $H=H_g$, which can have important
implications for the low-temperature surface resistance $R_s$ at high
RF fields $H(t)\sim H_c$. The BCS
surface resistance $R_s$ at small RF fields $H\ll H_c T/\Delta_0$, low frequencies $(\omega\ll \Delta_0)$ and
$T\ll T_c$ is given by \cite{mb,rs}
\begin{equation}
R_s =\omega^2\frac{A(\ell,\omega)}{T}\exp\left(-\frac{\Delta_0}{T}\right),
\label{rs}
\end{equation}
where the factor $A$ depends on the mean free path $\ell$ and (weakly) on the RF frequency $\omega$. The main Boltzmann factor $\exp(-\Delta_0/T)$ accounts for the exponentially small density of thermally-activated quasiparticles due to the zero density of states $N(\epsilon)$ for the energies $\epsilon<\Delta_0$. In the presence of Meissner current, the quasiparticle gap $\epsilon_g(J)$ shifts to smaller energies, giving rise to a highly nonlinear dependence of the surface resistance on the RF field amplitude at low temperatures:
\begin{equation}
R_s\propto \exp\left(-\frac{\epsilon_g(H)}{T}\right)
\label{rsh}
\end{equation}
As the field increases, $R_s(T,H)$ becomes essentially dependent on $H$ if the field-induced change of $\epsilon_g(H)$ is of the order of $T$. \cite{Gurevich06} In the clean limit for which $\epsilon_g(H)=(1-H/H_g)\Delta_0$, this condition takes the form
\begin{equation}
H\gtrsim TH_c/\Delta_0.
\label{ht}
\end{equation}
Therefore, the dc superheating field $H_s$ in the clean limit has
no direct relevance to the maximum RF magnetic field at which the Meissner state can exist.
Moreover, even at fields $H_\omega$ smaller than $H_g<H_c$, the RF field starts
generating quasiparticles as the gap $\epsilon_g(H)$ becomes smaller than
the RF frequency: $\epsilon_g(H_\omega)=\hbar\omega$. Hence
\begin{equation}
H_g -H_\omega \gtrsim \hbar\omega H_g/\Delta_0
\label{ho}
\end{equation}
As the magnitude of $H(t)$ approaches $H_\omega$, the surface resistance increases strongly, becoming of the order of $R_s(T)$ in the normal state. Calculation of the nonlinear surface impedance requires solving equations for the nonequilibrium Keldysh functions that take into account not only current pairbreaking, but also the effect of RF field on the quasiparticle distribution function determined by collisions of electrons with impurities and phonons \cite{Kopnin}. We will not discuss here this complex problem in detail, but only make a few qualitative remarks based on our solutions of the dc Eilenberger equations that capture the essential effect of impurities on the quasiparticle density of states at high RF fields.

Nonmagnetic impurities can strongly affect the field dependence of $R_s$ because they reduce $H_s$ by only a few percentages but restore the gap $\epsilon_g$ in the quasiparticle spectrum at $H=H_s$, as shown in Figs.\ref{fig11} and \ref{fig12}. In the dirty limit, $\alpha>1$, the gap $\epsilon_g\simeq 0.3\Delta_{00}$ may therefore be big enough to ensure both the exponentially small surface resistance in Eq. (\ref{rsh}) at low temperatures and the lack of quasiparticles generated by the RF field with $\omega < \epsilon_g$. As a result, nonmagnetic impurities can drastically reduce the field-induced increase of $R_s$ as compared to the clean limit. Moreover, the dc superheating field $H_s \approx 0.83 H_c$ at $\alpha > 1$ can now be regarded as a true maximum field amplitude at which the Meissner state can survive under low frequency RF fields. This conclusion may be essential for the  materials optimization for superconducting cavities used in particle accelerators.

\begin{acknowledgments}
%This work was supported by the US Department of Energy and the Argonne National Laboratory through the subcontract NSC99-2911-I-216-001.
Lin would like to thank Dr.~P.~Matlock for valuable discussions.
Funding for this work was provided by American Recovery and Reinvestment Act through the US Department of Energy, Office of High Energy Physics Department of Science to Argonne National Laboratory. NSC99-2911-I-216-001.
\end{acknowledgments}

%---------------------------------------------------------------bibliography

\end{document}